\documentclass[final,5p,twocolumn]{elsarticle}

\journal{PLB}

\begin{document}

\begin{frontmatter}


\title{Direct mass measurements of Cd isotopes show strong shell gap at N=82}
\author[GSI]{R.~Kn\"obel\corref{cor}}
\author[JLU]{M.~Diwisch\corref{PHD}}
\author[GSI]{F.~Bosch}
\author[GSI]{D.~Boutin}
\author[GSI]{L.~Chen}
\author[GSI]{C.~Dimopoulou}
\author[GSI]{A.~Dolinskii}
\author[GSI]{B.~Franczak}
\author[GSI]{B.~Franzke}
\author[GSI,JLU]{H.~Geissel}
\author[MSU]{M.~Hausmann}
\author[GSI]{C.~Kozhuharov}
\author[GSI]{J.~Kurcewicz}
\author[GSI]{S.~A.~Litvinov}
\author[TUD,GSI]{G.~Martinez-Pinedo}
\author[GSI]{M.~Mato\v{s}}
\author[GSI]{M.~Mazzocco}
\author[GSI]{G.~M\"{u}nzenberg}
\author[Saitama]{S.~Nakajima}
\author[GSI]{C.~Nociforo}
\author[GSI]{F.~Nolden}
\author[Niigata]{T.~Ohtsubo}
\author[Tsukuba]{A.~Ozawa}
\author[SOL]{Z.~Patyk}
\author[GSI,JLU]{W.~R.~Pla\ss}
\author[JLU,GSI]{C.~Scheidenberger}
\author[GSI]{J.~Stadlmann}
\author[GSI]{M.~Steck}
\author[CN,GSI]{B.~Sun}
\author[Saitama]{T.~Suzuki}
\author[USU]{P.~Walker}
\author[GSI]{H.~Weick}
\author[TUD]{M.-R.~Wu}
\author[GSI]{M.~Winkler}
\author[Saitama]{T.~Yamaguchi}
\cortext[cor]{Corresponding author}
\cortext[PHD]{Part of doctoral thesis at JLU Gie\ss en (2015)}
\address[GSI]{GSI Helmholtzzentrum f\"ur Schwerionenforschung GmbH, 64291 Darmstadt, Germany}
\address[JLU]{II.~Physikalisches Institut, Justus-Liebig-Universit\"at Gie\ss en, 35392 Gie\ss en, Germany}
\address[MSU]{Michigan State University, East Lansing, Michigan 48824, USA}
\address[TUD]{Physikalisches Institut, Technische Universit\"at Darmstadt, 64289 Darmstadt }
\address[Saitama]{Department of Physics, Saitama University, Saitama 338-8570, Japan}
\address[Niigata]{Department of Physics, Niigata University, Niigata 950-2181, Japan}
\address[Tsukuba] {Institute of Physics, University of Tsukuba, Ibaraki 305-8571, Japan}
\address[SOL]{National Centre for Nuclear Research - NCBJ Swierk, Ho$\dot{z}$a 69, 00-681 Warszawa, Poland}
\address[CN]{School of Physics and Nuclear Energy Engineering, Beihang University, Beijing 100191, China}
\address[USU]{Department of Physics, University of Surrey, Guildford, GU2 7XH, United Kingdom}

\begin{abstract}
A $^{238}$U projectile beam was used to create cadmium isotopes via abrasion-fission
at 410 MeV/u in a beryllium target at the entrance
of the in-flight separator FRS at GSI. The fission fragments were separated with the
FRS and injected into the isochronous storage ring ESR for mass measurements. The Isochronous Mass Spectrometry (IMS) was performed under two different experimental conditions, with and without B$\rho$-tagging at the dispersive central focal plane of the FRS. In the experiment with B$\rho$-tagging the magnetic rigidity of the injected fragments was determined by an accuracy of $2\cdot10^{-4}$. A new method of data analysis, using a correlation matrix for the combined data set from both experiments, has provided mass values for 25 different isotopes for the first time. The high selectivity and sensitivity
of the experiment and analysis has given access even to rare isotopes detected with a few atoms per week. In this letter we present for the $^{129,130,131}$Cd isotopes mass values directly measured for the first time. The Cd results clearly show a very pronounced shell effect at N=82 which is in agreement with the conclusion from $\gamma$-ray spectroscopy of $^{130}$Cd and confirms the assumptions of modern shell-model calculations.
\end{abstract}


\end{frontmatter}



\section{Introduction}

Accurate mass measurements reflect  details of the evolution of nuclear structure and stability as well as the energy levels and spatial distributions of the bound nucleons \cite{Bohr-Mottelson-1998}.
A first microscopic explanation of the observed shell structure and the corresponding magic numbers \cite{Goeppert-Mayer-1948,Haxel-1949} of neutrons and protons, at which the nuclei have larger
binding energies, refined the understanding of nuclear properties and manifested the
application of the shell model.
The advent and application of radioactive nuclear beam
facilities \cite{Geissel-Encyclopedia-2013} and novel mass spectrometers \cite{Litvinov-2013} have constantly enlarged the number of known isotopes with unusual proton-to-neutron ratios and thus explored the properties at the outskirts of the chart of nuclides and soon it became evident that the nuclear shell structure can change towards the driplines. Shell quenching, complete disappearance, or even new magic numbers have been theoretically predicted \cite{Dobaczewski.Hamamoto.ea:1994,Otsuka-PRL-2006} and observed in experiments \cite{Thibault-1975,Ozawa-2000,Kanungo-2013,Guillemaud-Mueller.Detraz.ea:1984,Motobayashi.Ikeda.ea:1995}.
The best known examples, both theoretically and
experimentally, are the $N=20$ and $N=28$ ``islands of
inversion''~\cite{Guillemaud-Mueller.Detraz.ea:1984,Motobayashi.Ikeda.ea:1995,Bastin.Grevy.ea:2007,Brown:2001,Caurier.Martinez-Pinedo.ea:2005,Caurier.Nowacki.Poves:2014}
where the gain in correlation energy driven by quadrupole deformation
is able to overcome the normal level population deduced from the standard spherical mean field. As a result the traditional $N=20$ and $N=28$ shell closures
completely disappear. It has also been argued that such a shell quenching  
would occur for neutron-rich N=82 nuclei. This phenomenon was originally
suggested based on Skyrme Hartree-Fock-Bogouliuvov
calculations~\cite{Dobaczewski.Hamamoto.ea:1994}. Almost
simultaneously it was realized that the occurrence of an abundance
trough around $A\sim 115$ in r-process
calculations~\cite{Kratz.Bitouzet.ea:1993} could be cured using a mass
model with a quenched shell gap far from
stability~\cite{Chen.Dobaczewski.ea:1995,Pearson.Nayak.Goriely:1996,Pfeiffer.Kratz.Thielemann:1997}. The
abundance trough around $A\sim 115$ is associated with a `saddle point
behavior' seen in the two-neutron separation energies for $Z\approx
40$ and $N=75$--82 in several mass models related to a transition from
deformed nuclei around $N\sim 75$ to spherical nuclei at
N=82~\cite{Grawe.Langanke.Martinez-Pinedo:2007}. In mass models with
a quenched shell-gap like the quenched extended Thomas-Fermi model
(ETFSI-Q)~\cite{Pearson.Nayak.Goriely:1996} the deformation is greatly
reduced and consequently the `saddle point behavior' in the
two-neutron separation energies disappears. However, it should be
pointed that the `saddle point behavior' and the quenching of the
shell gap are not necessarily
related~\cite{Grawe.Langanke.Martinez-Pinedo:2007} as the first could
also be associated to instabilities of mean-field models in regions of
shape coexistence that require the inclusion of beyond-mean field
correlations~\cite{Rodriguez.Arzhanov.Martinez-Pinedo:2015}. It is
interesting to notice that the `saddle point behavior' is absent in the
Duflo-Zuker shell-model inspired mass model~\cite{DZ-28}
and consequent does not result in an abundance trough around $A\sim
115$ despite of predicting a relatively large N=82 shell gap far
from stability. Furthermore, the strong abundance troughs predicted in
calculations were largely due to the use of the so called classical
r-process approximation which assumes an
$(n,\gamma) \leftrightarrow (\gamma,n)$ equilibrium and an
instantaneous freeze-out of neutron
captures~\cite{Kratz.Bitouzet.ea:1993}. Nowadays, we know that the
r-process occurs in a very dynamical environment with continuously
changing astrophysical conditions and that the competition between
neutron captures, beta-decays and fission has an  important impact in
the final
abundances~\cite{Arcones.Martinez-Pinedo:2011, Mendoza-Temis.Martinez-Pinedo.ea:2014}.

Given the relevance to both r-process nucleosynthesis and nuclear
structure of the existence of a new `island of inversion' approaching
$Z\sim 40$ for N=82, there have been many experimental attempts to
provide evidence for the quenching of the shell gap. It should be
clarified that quenching is sometimes used as a synonymous of
reduction of the shell gap. However, in the context of this letter we
will use `quenching' according to the most common usage of
disappearance of shell effects pointing to an `island of inversion'
analogous to the well-studied cases around $^{32}$Mg for $N=20$ and
$^{42}$Si for $N=28$. The experimental information about the evolution
of the N=82 gap is so far mainly indirect.

The present Isochronous Mass Spectrometry (IMS) experiment and the previous Penning trap mass
measurements for the tin isotopes~\cite{Dworschak.Audi.ea:2008} yield direct 
information on the shell effect. We have applied IMS to determine the masses of $^{129,130,131}$Cd isotopes. The cadmium and tin results provide a direct experimental evidence for the evolution of
the N=82 shell gap south of $^{132}$Sn.

\section{Experiment and data analysis}

A 410 MeV/u $^{238}$U projectile beam was extracted from the
synchrotron SIS-18 \cite{SIS} with an average intensity of
1$\cdot10^9$/spill and impinged on a 1032~mg/cm$^{2}$ beryllium
target at the entrance of the fragment separator FRS \cite{FRS}.
Neutron-rich fission fragments created via abrasion-fission were
separated in flight with the FRS applying pure  magnetic rigidity
(B$\rho$) separation with the standard ion-optical operation mode.
The separation mode, without degraders, was enabled by the large mean velocity
difference of the projectile fragments and fission products and
the restricted angular acceptance of the FRS. Practically this means, a suitable $B\rho$-selection with the FRS can provide fission-fragment beams without significant contributions of projectile fragments. The ions of interest
were injected into the Experimental Storage Ring ESR \cite{ESR}
for IMS \cite{Dolinskii-2007, FGM-Review} at a
mean velocity corresponding to the transition energy of
$\gamma_{t}$~=~1.41. The magnetic fields of the FRS and ESR were
set for $^{133,135,136}$Sn ions in different runs, i.e., these
isotopes were subsequently centered  at the optical axis.

The ESR was operated in the isochronous mode \cite{Hausmann-2000}
without application of any cooling. This means that the velocity
spread of the fragments is determined by the B$\rho$ acceptance of the ion-optical
system. In a previous publication \cite{Geissel-2006} we have demonstrated that for IMS experiments in addition to the revolution time of the stored ions the magnetic rigidity or velocity measurement is required, because the isochronicity is strictly  realized only for a  single mass-over-charge $(m/q)$ value.

In principle, this additional measurement is not required for Schottky Mass Spectrometry (SMS) because the relative velocity spread of the different stored and cooled ions can be as low as 10$^{-7}$. Nevertheless, our refined SMS analysis has revealed that an additional influence of the mean velocity causes an observed
correlation \cite{Chen-2012}.

The method of IMS including B$\rho$-tagging can be illustrated by the simple first-order formula

\begin{equation}\label{IMS}
\frac{d(m/q)}{m/q} = {\gamma}^2 \frac{d(T)}{T} + (1-\frac{{\gamma}^2}{{\gamma_t}^2})
\frac{dB\rho}{B\rho},
\end{equation}

where $T, \gamma_t$, and $ \gamma$ are the revolution time, the transition energy, and
the relativistic Lorentz factor, respectively.

The additional velocity ($v$) and magnetic-rigidity measurements in FRS-ESR IMS
experiments require special methods due to the ESR operation with fast extracted
ion bunches characterized by a width of $(0.2-0.5)~\mu$s. Particle detectors inside
the FRS would have severe problems to identify event-by-event the fragments and accurately measure $v$ and $B\rho$. A B$\rho$-resolution of $10^{-4}$ or better is required to achieve a mass resolution of about 200 keV for m/q close to ideal isochronicity, details are presented in reference \cite{Diwisch-PHD}. In this context, one has to take into account that the FRS transmission is $d(B\rho)/(B\rho)= 2\%$ and the corresponding 
ESR injection acceptance is roughly one order of magnitude less. Therefore, mechanical
slits with a slit opening of $\pm$0.5 mm placed at the central dispersive focal plane
of the FRS were used in a pilot IMS experiment \cite{Sun:2008}. The slits defined in this way the magnetic rigidity (B$\rho$-tagging) of each injected ion with an accuracy of $2\cdot10^{-4}$.
The operation of the new isochronous Rare RI-Ring at RIKEN \cite{Ozawa-2012} can easily implement additional B$\rho$ and $v$ measurements event-by-event, because of the DC beam from the cyclotron.
IMS measurements have been performed with and without B$\rho$-tagging for the same settings of the magnetic fields of the FRS and ESR. The revolution time of the circulating
ions in the ring were measured with a time-of-flight (ToF) detector equipped with a thin carbon foil and two micro-channel-plate (MCP) branches \cite{Hausmann-2000} placed in a homogeneous magnetic dipole field of about 8.4~mT. The created secondary electrons in the carbon foil were isochronously
deflected onto the MCPs to generate timing signals at each turn. The signals were recorded
with commercial digital oscilloscopes (Tektronix TDS 6154C,  40~GS/s, 15~GHz; LeCroy LC584AM,  4~GS/s, 1~GHz)

The data sets of the two different experiments, with and without B$\rho$-tagging, were combined and analysed with the new correlation-matrix method \cite{Radon-2000,Litvinov-2004}. In this work we used for the first time a variable correlation factor dependent on the measured $\frac{m}{q}$ . 
A detailed description of this novel analysis will be presented
in a forthcoming publication \cite{Knoebel-2015}. The advantage of this new analysis
is that we could include ions with very low statistics down to a few events of a single isotope. In this way we can now present more than 20 new mass values which were not included in previous IMS evaluations of the same experiments \cite{Matos:2004, Sun:2008}.
\\

\section{Results and discussion}

\begin{figure}[tb]
\centering\includegraphics[width=88mm]{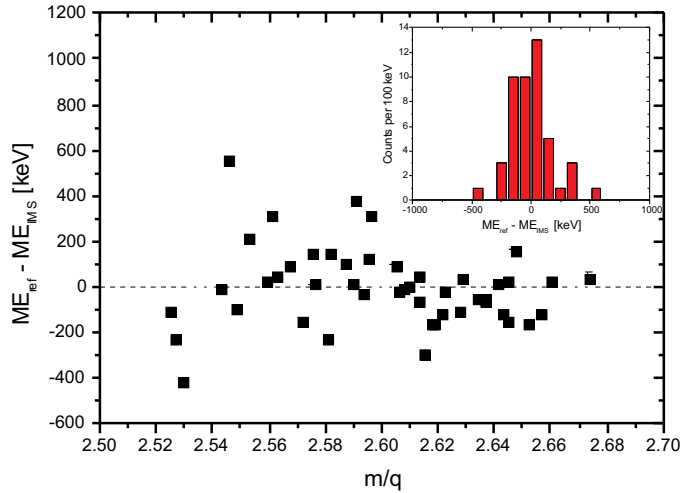}
\caption{The systematic error of our mass measurement was determined by reanalysis of each reference mass, i.e.,
 each reference mass is treated subsequently as an unknown species. The distribution of this analysis is depicted as a function of mass-over-charge values and shown in the insert as a projected histogram.}
\label{Systematic-errors}
\end{figure}

The first check of the reliability of the new data analysis is the determination of the systematic error of the combined experiments. An average of 1600 events per reference mass could be applied in the analysis with the combined data sets. We determined the systematic error by using all reference masses and subsequently treat each of these masses as being unknown. 
The procedure can be illustrated by equation \ref{syst-error} and figure
\ref{Systematic-errors}.

\begin{equation}\label{syst-error}
\sum_i^{n} \frac{(M_i-M_i^{ref})^2}{(\sigma   _i^{ref})^2+(\sigma_i^{stat})^2+(\sigma^{syst})^2}=N_n,
\end{equation}
where $ M_i^{ref}$ are the mass values and $\sigma_i^{ref}$ the uncertainties of the reference nuclides. $\sigma_i^{stat}$ are the statistical errors of the measured masses $M_i$. $N_{n}$ is the number of reference masses and $\sigma^{syst}$ the systematic error. In this analysis 47 reference masses have been used \cite{JYFL12}.
The investigation covers the full m/q range of the present analysis. The mean value of the projected distribution is 1.29~keV. The deduced systematic error  is 172~keV (standard deviation).
For most of the new mass values the systematic error is the dominant contribution to the total error which results from the sum of the variances.
The statistical error is determined by the uncertainty of the revolution time, which strongly depends on the number of turns for which the timestamps have been recorded.
In the matrix method the statistical error for each nuclide was calculated by the square root of the diagonal elements of the inverse matrix. Details will be presented in reference \cite{Knoebel-2015}.
\\ 
A main motiation of the present mass measurements is the investigation the N=82 shell gap for Cd isotopes.
We briefly review previous experiments and conclusions before presenting our results. Ref.~\cite{Dillmann-2003} determined the beta-decay half-live of the $^{130}$Cd nuclide and its $Q_{\beta}$ value. The mass of $^{130}$Cd in the Atomic Mass Evaluation 2012 (AME12)~\cite{AME12} is based on this $Q_{\beta}$ value. It was concluded that the large measured $Q_{\beta}$-value of $8.32$~MeV is an indication of shell quenching because the experimental value was in good agreement with the predictions of the ETFSI-Q mass model \cite{ETFSI-Q}. However, it is about 1~MeV larger than predicted by other mass models such as the Finite Range Droplet Model~\cite{FRDM} or the Duflo-Zuker model~\cite{DZ-28}. A recent experiment \cite{Lorusso.Nishimura.ea:2015} has determined a substantially shorter beta-decay half-live with more statistics for the $^{130}$Cd nuclides, which questions the determination of the $Q_{\beta}$-value in \cite{Dillmann-2003}. The  measurement of the $2^+$ excitation energy for $^{128}$Cd showed that their value was the smallest of the known N=80 isotones~\cite{Kautzsch.Walters.ea:2000}. Again, it was interpreted as evidence of shell quenching. However, beyond-mean field calculations~\cite{Rodriguez.Egido.Jungclaus:2008} showed that such a low excitation energy could be explained by the presence of prolate configurations in the wave function of $^{128}$Cd. This may explain the measured small magnetic moment observed in $^{127}$Cd~\cite{Yordanov.Balabanski.ea:2013}. The same calculations predicted a much larger excitation energy of the $2^+$ level in $^{130}$Cd. This excitation energy has been measured to be 1.321~MeV \cite{Jungclaus-2007} in agreement with shell-model calculations that assumed a N=82 shell closure. This fact together with the large excitation energy of the $2^+$ state in $^{126}$Pd~\cite{Watanabe-2013} (1.311~MeV) has been interpreted as evidence for a robust N=82 shell closure for both cadmiun and palladium isotopes. Recently, a $\gamma$-spectroscopy experiment \cite{Taprogge-2014} has determined the $3/2^-$ state in $^{132}$In corresponding to the $p_{3/2}$ proton-hole state in $^{132}$Sn. It is interesting to note that these new data are in perfect agreement with the predictions of the Duflo-Zuker monopole hamiltonian~\cite{Duflo.Zuker:1999} which is the base for the Duflo-Zuker mass model. The new experimental data has allowed to determine all relevant proton single particle states showing that no proton subshell closures are expected for light N=82 isotones pointing to a significant reduction of the N=82 shell gap in analogy to the reduction observed in isotones heavier than N=82. This reduction does not necessarily mean a quenching of the shell gap. It is well known that shell gaps reach a maximum value for magic numbers of protons and neutrons \cite{Zeldes.Dumitrescu.Koehler:1983}.
\\

\begin{figure}[t]
\centering\includegraphics[width=0.48\textwidth]{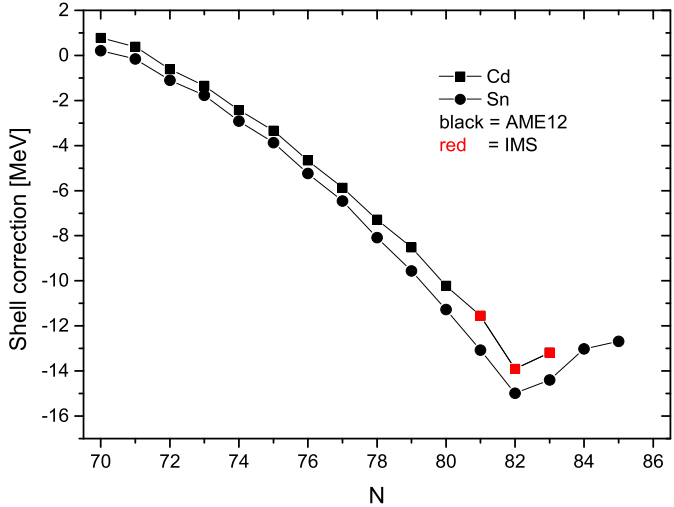}
\caption{Experimental shell effect. Difference of the measured
mass values and the smooth Weizs\"acker formula given by Eq. \ref{BW}  for the elements Sn and Cd. The parameters used in the formula are
presented in Table \ref{LQD-parameters}. The curves clearly show the extra binding energy due to the contribution of shell structure for the
Sn and Cd nuclei in this mass region of N=82.}
\label{AME12IMS-BW}
\end{figure}

Figure~\ref{AME12IMS-BW} shows the difference of the experimental mass
values for Sn and Cd isotopes and the prediction of the
liquid drop model \cite{Bohr-Mottelson-1998,Weizsäcker-1935}. The
liquid drop parameters have been deduced from a fit
to the tabulated values of the Atomic Mass Evaluation 2012 (AME12)
\cite{AME12}.

The liquid-drop binding energy $B_{LD}$ has the form
	\[B_{LD}=b_{vol}A -b_{surf}A^{2/3}-\frac{1}{2}
b_{sym}\frac{(N-Z)^2}{A}-
\]
\begin{equation}\label{BW}
 \frac{3}{5}\frac{Z^2e^2}{R_c} - \left\{\begin{array}{cl} 0, & \mbox{e-e nuclei}\\ b_{pair}A^{-1/2}, & \mbox{e-o or o-e nuclei}\\ 2\cdot b_{pair}A^{-1/2}, & \mbox{o-o nuclei}\end{array}\right.
\end{equation}
with $R_c=1.24$ fm $\cdot A^{1/3}$. 
\\
\begin{table}[t!] 
\caption{Fitted liquid-drop parameters. The parameters, in units of MeV,
for the Weizs\"acker formula have been obtained by a fit to the
mass values of reference \cite{AME12}.}

\begin{center}
\begin{tabular}{|c|c|c|c|}
  \hline
 $b_{vol}$ & $b_{surf}$ & $b_{sym}$ & $b_{pair}$ \\ \hline
 15.747 & 17.603 & 47.494 & 12.822  \\
\hline
\end{tabular}
\end{center}
\label{LQD-parameters}
\end{table}
\begin{figure}[h]
\centering\includegraphics[width=88mm]{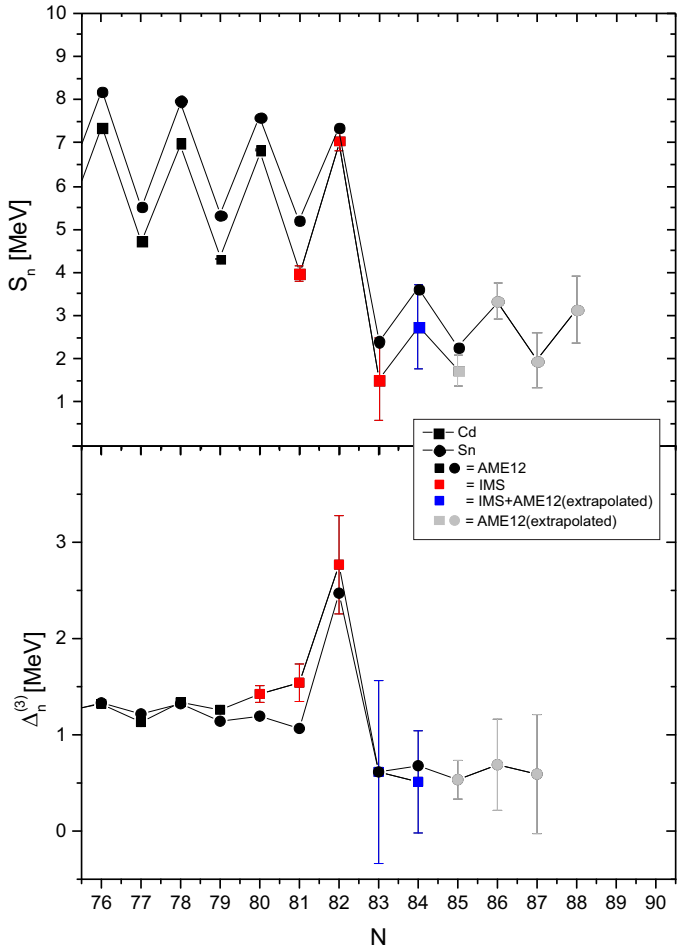}
\caption{In the upper panel experimental one-neutron separation energies ($S_n$) are shown for Cd and Sn isotopes. In the lower panel the corresponding pairing-gap energies are presented.}
\label{Sn_values}
\end{figure}

In figure \ref{AME12IMS-BW} the differences between experimental mass excess (ME) and the corresponding liquid-drop values are plotted. The ME is the difference between the atomic
mass and the corresponding mass number, both expressed in atomic mass units. The liquid-drop values are calculated with equation \ref{BW} and the parameters in table \ref{LQD-parameters}. The strong extra binding
energy  for the neutron-rich Cd isotopes is clearly observed at
N=82, it amounts to roughly -14 MeV. The error bars for
$^{129,130}$Cd are well within the symbol size whereas for
$^{131}$Cd we have a large statistical error because only two
events have been recorded with few number of turns.
The experimental results in figure \ref{AME12IMS-BW} clearly demonstrate that the tin and cadmium isotopes are characterized by the same difference of shell corrections at N=82.
This observation contradicts reference \cite{Dillmann-2003} and clarifies the situation
of the different statements in references \cite{Jungclaus-2007,Gorska-2009,Taprogge-2014}.

The experimental ME values for $^{129,130,131}$Cd nuclei
and the errors are presented in table \ref{Cd-masses}.
Our directly measured mass value for $^{130}$Cd atoms is 600 keV lower than the value
deduced from $Q_\beta$ measurements \cite{Dillmann-2003}. This difference could be caused by a contribution of an excited state in the $\beta$-decay experiment. In principle, also in our measurements for $^{129,131}$Cd nuclei unresolved isomers could contribute \cite{Taprogge-2014,Taprogge-2014b} whereas in $^{130}$Cd the lifetime is too short to interfere in this experiment. However, in this case the observed strong shell effect for Cd isotopes at N=82 would be even stronger for pure ground-state masses. The nearly equal shell effect for Sn and Cd is a striking feature of the present experimental results.

 Other established results from mass measurements to discuss the evolution of nuclear
structure and shells are the nucleon separation energies and their
derivatives. The one-neutron separation energy $S_n$ and the pairing gap energy $\Delta_n^{(3)}$ is defined by:
\begin{equation}\label{S_n}
S_n = -M(Z,N) + M(Z,N-1) + M(0,1)\\
\end{equation}
\begin{equation}\label{Delta3n}
\Delta_n^{(3)}=(-1)^N\frac{1}{2}\left[ S_n(Z,N)-S_n(Z,N+1)\right)]\\
\end{equation}
A and Z are the  mass and proton numbers, respectively.

\begin{table}[h!] 
\caption{Measured mass excess values ($ME$) of Cd isotopes. The statistical ($\sigma_{stat}$)  and the total ($\sigma_{total}$) errors
 are tabulated. The systematic error is 172 keV for all the new masses measured. In addition, the extrapolated ($\#$) values \cite{AME12} and the $^{130}$Cd mass deduced from $Q_{\beta}$-measurement \cite{Dillmann-2003} are listed.}
\begin{center}
\begin{tabular}{|c|c|c|c|c|c|}
  \hline
 isotope & $ME$ & $\sigma _{stat}$ & $\sigma _{total}$  & $ME_{AME12}$ & counts \\
     & $[keV]$ & $[keV]$ & $[keV]$ & $[keV]$ &\\   \hline
  $^{129}$Cd & -63145 & 17 & 173 &  -63509(196)$\#$& 18 \\ 
  $^{130}$Cd & -62131 & 41 & 177 &  -61534(164)& 5\\ 
  $^{131}$Cd  & -55583 & 925 & 941 &  -55331(196)$\#$& 2\\
  \hline
\end{tabular}
\end{center}
\label{Cd-masses}
\end{table}

In the upper panel of figure \ref{Sn_values}, the shell evolution is
manifested by our experimental (indicated in red) or AME12 one-neutron separation
energies and their derivative near N=82. In the lower panel, we
display an estimation of the gap representing the single particle spectrum (3-mass formula) which in principle is the derivative of
the S$_n$-values. Both presentations demonstrate a strong shell
effect at N=82 for Sn and Cd isotopes in complete agreement with
the characteristics of figure \ref{AME12IMS-BW}.
Again it is a striking experimental observation that Sn and Cd isotopes are governed by the same strong shell gap at N=82. The magnitude of the shell gap is within the experimental uncertainties the same.



\begin{figure}[h]
\centering\includegraphics[width=105mm]{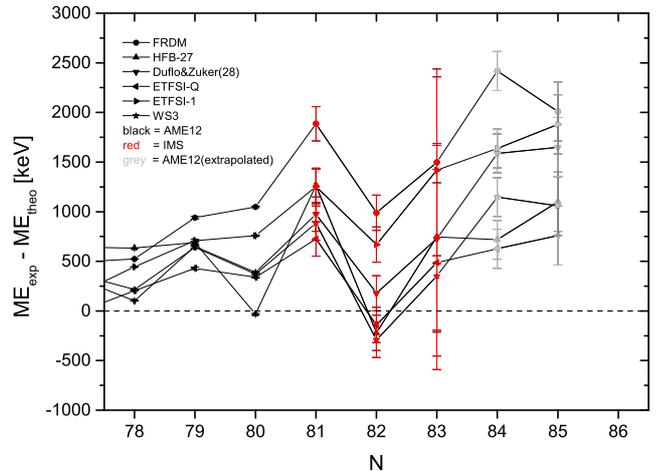}
\caption{Comparison of experimental mass values for Cd with different theoretical predictions}
\label{Theories}
\end{figure}

A comparison of the experimental values for Cd isotopes with different theoretical mass models is presented in figure \ref{Theories}. The models are based on microscopic-macroscopic descriptions \cite{FRDM,WS3,ETFSI-1,ETFSI-Q},
the Hartree-Fock-Bogoliubov (HFB) theory \cite{HFB-27}, and the shell-model inspired model of Duflo-Zuker~\cite{DZ-28}.

In general, the comparison with theoretical predictions clearly
demonstrates that the agreement with experimental results is
getting worse towards the shell closure at N=82 and of course
in the n-rich region, where no experimental results were available
before. There are basically 3 different regions in this comparison. In the first region for N$\leq$81 the theories are relatively similar in their trend but all predict too low mass values (over-binding). At the shell the predictions are quite unreliable and beyond N$>(82)$, where no experimental data exists, they widely scatter. The upper panel show the described comparison for Cd isotopes and the lower panel the corresponding results for Sn isotopes. The masses of the presented Sn isotopes have been measured with Penning traps \ with higher resolving power and should not have influenced by unresolved isomeric states. The predictive power of the different models can be quantitatively characterized by the $\sigma_{rms}$ values. They are
listed in table \ref{rms-values} for Sn and Cd isotopes in the experimentally known range of AME12 and
for Cd isotopes presented in figure \ref{Theories}.

\begin{table}[h!] 
\caption{Comparison of measured data with models. RMS deviations for Sn and Cd  isotopes for  different theoretical
models are presented. In this comparison the new IMS values and the tabulated experimental values of AME12 are included.
$\sigma_{rms,1} $,$\sigma_{rms,2} $ represent the values for the cases of all experimentally known Cd isotopes and the Cd isotopes in the mass range as shown in figure 4, respectively. 
}
\begin{center}
\begin{tabular}{|c|c|c|}
  \hline
	&	  	  $\sigma_{rms,1}$	&	  $\sigma_{rms,2}$	\\	
	&	$ [keV]$	&	$ [keV]$		\\	\hline
FRDM \cite{FRDM} 		&	619	&	1537	\\	
 HFB-27 \cite{HFB-27}		&	523	&		775	\\	
 DZ28 \cite{DZ-28}		&	315	&		958	\\	
 ETFSI-Q \cite{ETFSI-Q}	&	342	&		512	\\	
 ETFSI-1 \cite{ETFSI-1}	&	479	&		1200	\\	
 WS3 \cite{WS3}	&	396	&		709	\\	\hline

\end{tabular}
\end{center}
\label{rms-values}
\end{table}

\begin{figure}[htb]
  \centering
  \includegraphics[width=\linewidth]{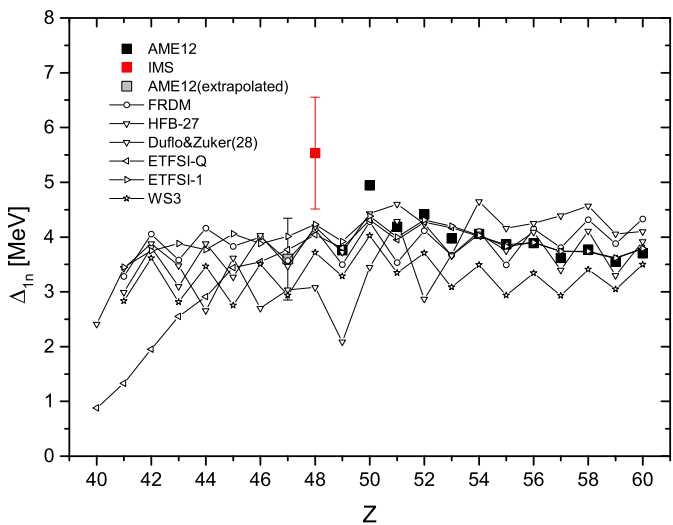}
  \caption{One-neutron gap (see text) for $N=82$ isotones. \label{fig:gaps}}
\end{figure}

The comparison with theories over a larger Z range 
for $N=82$ isotones is presented in figure~\ref{fig:gaps} for the one-neutron
gap, $\Delta_{1n}(Z,82)=S_n(Z,82)-S_n(Z,83)$. Experimental $S_n$ values deduced from mass measurements exist only above cadmium.
The figure clearly shows the
large gap measured for $^{130}$Cd that is within the experimental
uncertainties of similar magnitude than the one for $^{132}$Sn. No
theoretical model is able to reproduce such a large gap for Cd and Sn isotopes.
Looking at
the evolution of the one-neutron gap and averaging the odd-even staggering 
one observes that only the ETFSI-Q and HFB27 models predict a 
strong reduction towards Z=40.

The new data for Cadmium isotopes is not consistent with an
early quenching of the $N=82$ shell gap as claimed before by several
publications~\cite{Dillmann-2003,Kautzsch.Walters.ea:2000}. However,
it cannot rule out a disappearance of the shell gap for lighter
isotones. Therefore, it will be necessary to have experimental access to
isotones with $Z\leq 42$ for which the differences between models
become substantial. 


\section{Summary and Outlook}

IMS experiments have been performed with and without B$\rho$-tagging at the FRS-ESR facilities at GSI.
A new method of data analysis using the correlation matrix for the combined data
of both types of experiments provided 25 new mass values in range of Ge to Ce (A=86-154) even for isotopes measured with very low statistics of a few atoms per week. The latter condition clearly demonstrates the high sensitivity and selectivity of the experimental method. In this letter we have presented the masses of the $^{129,130,131}$Cd isotopes which have been directly measured for the first time. The goal was to investigate the evolution of the shell gap of Cd isotopes compared to Sn isotopes at N=82. The Cd results clearly show a very pronounced shell effect at N=82 which is assumed by modern shell-model calculations. The experimental values for the shell corrections are roughly the same for Sn and Cd isotopes.

The goal of future IMS experiments will be to measure the lower elements ($40<Z<48$) at and near the N=82  shell. Only the mass values of the lower elements will be decisive for the observation of possible shell quenching as it is proposed in several theoretical models. However, experimentally this interesting region can only be accessed by the next generation exotic nuclear beam facilities \cite{FAIR, RIBF, FRIB, EURISOL}.

\section*{Acknowledgements}

This work was supported by the German Federal Ministry of Education
and Research (BMBF) numbers 06GI9115I and 06DA7047I, the Helmholtz
Association (HGF) through the Nuclear Astrophysics Virtual Institute
(VH-VI-417) and the Helmholtz International Center for FAIR (HIC for
FAIR) within the framework of the LOEWE program launched by the State of
Hesse.






\end{document}